\documentclass[epj]{svjour}

\usepackage{graphicx}
\usepackage{fancyhdr}
\def\makeheadbox{{
 }}
\newcommand\ReTilde{\widetilde{\rm Re}}

\def\mytitle{My title}
\def\myauthors{My name}
\def\mytype{My type of session}
\def\mysession{My session}

\def\mytitle{One-loop corrections to three-body leptonic chargino decays} 
\def\myauthors{Krzysztof Rolbiecki}    
\def\mytype{Contributed Talk}
\def\mysession{Colliders - SUSY Phenomenology}

\begin{document}
\title{One-loop corrections to three-body leptonic chargino decays}
\author{Krzysztof Rolbiecki
\thanks{\emph{E-mail:} krzysztof.rolbiecki@fuw.edu.pl}%
}                     
\institute{Institute of Theoretical Physics, University of Warsaw,
Ho\.{z}a 69, 00-681 Warsaw, Poland}
%
\date{}
\abstract{We calculate full one-loop corrections to the genuine
three-body decays of the light chargino $\tilde{\chi}_1^\pm \to
\tilde{\chi}^0_1 \ell^\pm \nu_\ell$ in the Minimal Supersymmetric
Standard Model. We find that the corrections to the decay width can
be of the order of a few percent. We show also how radiative
corrections affect energy and angular distributions of the final
lepton.
} 
\maketitle
\section{Introduction} \label{intro}
\sloppypar Supersymmetry (SUSY)~\cite{susy} is one of the most
promising and best motivated extensions of the Standard Model (SM)
of particle physics. The search for SUSY is one of the main goals at
the present and future colliders. All SUSY theories contain
charginos, the spin-1/2 superpartners of charged gauge bosons and
Higgs bosons. In many scenarios charginos are expected to be light
enough to be copiously produced at future high energy colliders ---
the Large Hadron Collider (LHC)~\cite{LHC} and the International
Linear Collider (ILC)~\cite{ILC}.

Once the supersymmetric particles will have been discovered it will
be crucial to measure their masses, mixing, couplings and CP
violating phases to reconstruct the fundamental SUSY parameters and
get insight of physics at very high energy scales. To meet the
requirements of very high experimental precision at the ILC it is
very important to include in theoretical calculations higher-order
loop corrections to physical processes and observables.

Chargino production has been thoroughly analyzed at one-loop level
in the literature~\cite{fritzsche,oeller,tania}. Recently full
one-loop analysis of chargino decays was
publised~\cite{Fujimoto:2007bn}, however showing only corrections to
the decay widths and branching fractions of charginos.

In this note we report on the calculation of the full one-loop
corrections to the genuine three-body chargino decays to the
lightest neutralino, lepton and neutrino
\begin{eqnarray}
\tilde{\chi}_1^\pm \to \tilde{\chi}^0_1 \ell^\pm \!
\stackrel{_{_{(-)}}}{\nu_\ell}\: .\label{eq:decay}
\end{eqnarray}
The calculation was performed for the complex Minimal Supersymmetric
Standard Model (MSSM) and it allows inclusion of CP-violating
effects in future. In the presented results we include corrections
to the decay widths and to the energy and angular distributions of
the final lepton. We also show the impact of QED corrections. We
compare differences between the decay to electron and to $\tau$.

The paper is organized as follows. In Sec.~\ref{sec:2} we
recapitulate chargino and neutralino sectors of MSSM at the
tree-level. In Sec.~\ref{sec:3} we introduce renormalization scheme
and analyze the structure of one-loop corrections to the
decay~(\ref{eq:decay}). In Sec.~\ref{sec:4} we present our numerical
results and finally in Sec.~\ref{sec:5} we summarize our findings
and give outlook for future developments.

\section{Gaugino/higgsino sector of the MSSM} \label{sec:2}
\subsection{Chargino mixing}\label{sec:2.1}
In the MSSM, the tree-level mass matrix of the spin-1/2 partners of
the charged gauge and Higgs bosons, $\tilde{W}^+$ and $\tilde{H}^+$,
takes the form
\begin{eqnarray}
{\cal M}_C=\left(\begin{array}{cc}
  M_2       &   \sqrt{2}  m_W\sin\beta \\[2mm]
   \sqrt{2} m_W \cos\beta &     \mu
                  \end{array}\right)\: ,
\label{eq:massmatrix}
\end{eqnarray}
where $M_2$ is the SU(2) gaugino mass, $\mu$ is the higgsino mass
parameter, and $\tan\beta$ is the ratio $v_2/v_1$ of the vacuum
expectation values of the two neutral Higgs fields. By
reparametrization of the fields, $M_2$ can be taken real and
positive, while $\mu$ can be complex $\mu=|\mu|\,\,{\rm
e}^{i\Phi_\mu}$. Since the chargino mass matrix ${\cal M}_C$ is not
symmetric, two different unitary matrices are needed to diagonalize
it
\begin{eqnarray}
    U^* {\cal M}_C V^\dag =
    \left( \begin{array}{cc}
    m_{\tilde{\chi}^\pm_1} & 0\\
    0 & m_{\tilde{\chi}^\pm_2} \\
    \end{array}\right)\: .
\end{eqnarray}
$U$ and $V$ matrices act on the left- and right-chiral
$\psi_{L,R}=(\tilde{W},\tilde{H})_{L,R}$ two-component states
\begin{eqnarray}
\tilde{\chi}^R_j=U_{jk} \psi_k^R, \quad \tilde{\chi}^L_j=V_{jk}
\psi_k^L\: ,
\end{eqnarray}
giving two mass eigenstates $\tilde{\chi}^\pm_1$,
$\tilde{\chi}^\pm_2$.

\subsection{Neutralino mixing}\label{sec:2.2}
In the MSSM, four neutralinos $\tilde{\chi}^0_i$ ($i=1,2,3,4$) are
mixtures of the neutral U(1) and SU(2) gauginos, $\tilde{B}$ and
$\tilde{W}^3$, and the SU(2) higgsinos, $\tilde{H}^0_1$ and
$\tilde{H}^0_2$. The neutralino mass matrix in the $(\tilde{B},
\tilde{W}^3, \tilde{H}^0_1, \tilde{H}^0_2)$ basis
\begin{equation}
{\small    {\cal M}_N=
    \left( \begin{array}{cccc}
    M_1 & 0 & -m_Z c_\beta s_W & m_Z s_\beta s_W \\
    0 & M_2 & m_Z c_\beta c_W & -m_Z s_\beta c_W \\
    -m_Z c_\beta s_W & m_Z c_\beta c_W & 0 & -\mu \\
    m_Z s_\beta s_W & -m_Z s_\beta c_W & -\mu & 0 \\
    \end{array} \right)}
\end{equation}
is built up by the fundamental SUSY parameters: the U(1) and SU(2)
gaugino masses $M_1$ and $M_2$, the higgsino mass parameter $\mu$,
and $\tan\beta$ ($c_\beta = \cos\beta$, $s_W = \sin\theta_W$ etc.).
In addition to the $\mu$ parameter a non-trivial CP phase can also
be attributed to the $M_1$ parameter. Since the matrix ${\cal M}_N$
is symmetric, one unitary matrix $N$ is sufficient to rotate the
gauge eigenstate basis $(\tilde{B}, \tilde{W}^3, \tilde{H}^0_1,
\tilde{H}^0_2)$ to the mass eigenstate basis of the Majorana fields
$\tilde{\chi}^0_i$
\begin{eqnarray}
{\cal M}_{\rm diag} = N^*{\cal M}_N N^\dagger\: .
\end{eqnarray}
The mass eigenvalues $m_i$ ($i=1,2,3,4$) in ${\cal M}_{\rm diag}$
can be chosen real and positive by a suitable definition of the
unitary matrix $N$.

\subsection{Chargino decays at tree level}\label{sec:2.3}
\begin{figure}[!t]
\includegraphics[width=0.48\textwidth]{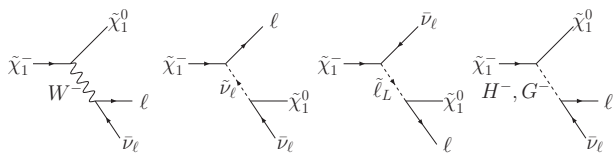}
\caption{Diagrams contributing to the leptonic three-body chargino
decay $\tilde{\chi}_1^\pm \to \tilde{\chi}^0_1 \ell^\pm \!
\stackrel{_{_{(-)}}}{\nu_\ell}$.} \label{fig:decay}
\end{figure}

At the tree level several channels contribute to the chargino decays
to leptons~(\ref{eq:decay}). This can be $W^\pm$, $\tilde{\ell}_L$,
$\tilde{\nu}_\ell$, $H^\pm$ and $G^\pm$ exchanges, see
Fig.~\ref{fig:decay}. For the decays to light fermions the Higgs and
Goldstone boson exchange channels can be neglected because the
contribution is strongly suppressed by the tiny Yukawa couplings.

\section{One-loop corrections}\label{sec:3}
\subsection{Renormalization scheme}\label{sec:scheme}
Since one-loop corrections introduce ultraviolet divergences we need
to apply a proper renormalization scheme to obtain physically
meaningful results. In this analysis we choose to work in the
on-shell scheme. This means that our renormalization conditions are
defined requiring the pole of the propagator and residue equal 1 at
the physical masses of particles. To regularize one-loop integrals
we use dimensional reduction, which preserves
supersymmetry~\cite{siegel}. For renormalization of SM parameters
and fields we follow closely procedure given in~\cite{denner}.

Renormalization of the chargino and neutralino sectors is performed
in the mass eigenstate basis~\cite{oeller}. We introduce in the
Langragian the wave function and mass counterterms with the
following substitution
\begin{eqnarray}
&&\tilde{\chi}_i \to (\delta_{ij} + \frac{1}{2} \delta
\tilde{Z}^L_{ij} P_L + \frac{1}{2} \delta \tilde{Z}^R_{ij} P_R)
\tilde{\chi}_j\: ,\nonumber \\
&&m_{\tilde{\chi}_i} \to m_{\tilde{\chi}_i} + \delta
m_{\tilde{\chi}_i}\: ,
\end{eqnarray}
where $\delta \tilde{Z}$ stands for either the chargino or the
neutralino field renormalization constants, $\delta \tilde{Z}^\pm$
or $\delta \tilde{Z}^0$, respectively. Similar substitution has to
be done for sleptons
\begin{eqnarray}
&& \left( \begin{array}{c} \tilde{f}_1 \\ \tilde{f}_2 \end{array}
\right) \to
    \left( \begin{array}{cc} 1+\frac{1}{2}\delta Z^{\tilde{f}}_{11} &
    \frac{1}{2} \delta Z^{\tilde{f}}_{12} \\ \frac{1}{2} \delta Z^{\tilde{f}}_{21} &
    1+\frac{1}{2}\delta Z^{\tilde{f}}_{22} \end{array} \right) \left(
    \begin{array}{c} \tilde{f}_1 \\ \tilde{f}_2 \end{array}
    \right)\: ,\nonumber \\
    && m^2_{\tilde{f}_i} \to m^2_{\tilde{f}_i} + \delta
    m^2_{\tilde{f}_i}\: .
\end{eqnarray}

For the renormalization of $\tan\beta$ we take the condition that
the CP-odd Higgs boson $A^0$ does not mix with $Z$ boson on-shell
\begin{eqnarray}
\frac{\delta\tan\beta}{\tan\beta} = \frac{1}{m_Z \sin 2\beta}
\mathrm{Im}\left[ \ReTilde\, \Sigma_{A^0 Z}(m_{A^0}^2)\right]\: ,
\end{eqnarray}
where $\Sigma_{A^0 Z}(m_{A^0}^2)$ is the self-energy for $A^0-Z$
mixing~\cite{Dabelstein:1994hb}.

We also have to define the chargino and neutralino rotation matrices
at one-loop level. We define them in such a way that they remain
unitary. Thus for charginos we have
\begin{eqnarray}
&&\delta U_{ij} = \frac{1}{4} \sum^2_{k=1} \Big(\delta
\tilde{Z}^{\pm,R}_{ik} - (\delta \tilde{Z}^{\pm,R}_{ki})^* \Big)
U_{kj}\: ,\nonumber \\
&&\delta V_{ij} = \frac{1}{4} \sum^2_{k=1} \Big(\delta
\tilde{Z}^{\pm,L}_{ik} - (\delta \tilde{Z}^{\pm,L}_{ki})^* \Big)
V_{kj}\: ,
\end{eqnarray}
and for neutralinos
\begin{eqnarray}
\delta N_{ij} = \frac{1}{4} \sum_{k=1}^4 \Big(\delta
\tilde{Z}^{0,L}_{ik} - \delta \tilde{Z}^{0,R}_{ki} \Big) N_{kj}\: .
\end{eqnarray}
Our procedure is kept general to accommodate complex phases that may
appear in the MSSM Lagrangian and to calculate CP-odd effects.

\subsection{Calculation at one loop}
\begin{figure}
\includegraphics[width=0.48\textwidth]{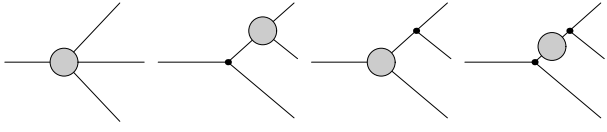}
\caption{Generic one-loop diagrams for chargino decays: box, vertex
and self-energy corrections.} \label{fig:generic}
\end{figure}

Radiative corrections to the chargino decay include the following
generic one-loop Feynman diagrams:  the box diagram contributions,
the virtual vertex corrections, the self-energy corrections. They
are displayed in Fig.~\ref{fig:generic}. To obtain finite results we
also have to include proper counterterms at vertices and
propagators.

Since the number of diagrams exceeds 200 an automatized computation
package has to be used. Generation and calculation of one-loop
graphs was performed using \texttt{FeynArts} \texttt{3.2} and
\texttt{FormCalc} \texttt{5.2} packages \cite{feynarts}. For
numerical evaluation of loop integrals we have used
\texttt{LoopTools} \texttt{2.2} \cite{looptools}. For our purpose we
have included in \texttt{FeynArts} model file for MSSM the necessary
counterterms derived as it was described in Sec.~\ref{sec:scheme}.
All results have been checked against UV-finiteness.

\subsection{QED corrections}
\begin{figure}
\includegraphics[width=0.48\textwidth,height=0.15\textwidth,angle=0]{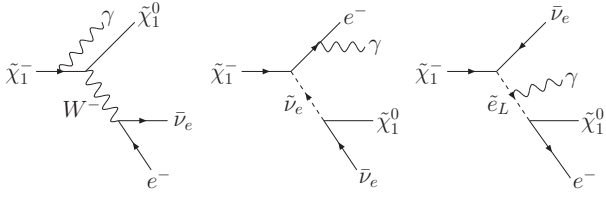}
\caption{Sample diagrams for photonic corrections to chargino
decays.} \label{fig:photonic}
\end{figure}

Some of the one-loop diagrams contain virtual photon exchange. Since
photon is a massless particle this leads to infrared divergences in
one-loop integrals. They have to be regularized using unphysical,
finite photon mass. These contributions cannot be separated from the
weak corrections in a gauge invariant and UV finite way. To obtain
physically meaningful results one has to include photon emission
from charged particles appearing at the tree-level. Sample diagrams
have been depicted in Fig.~\ref{fig:photonic}.

To cancel IR divergences it is enough to include soft photon
emission, i.e.\ emission of photons with energy $E_\gamma \leq
\Delta E$, where $\Delta E$ is small compared to the energy scale of
the process. However, this procedure gives us the result which
depends on an unphysical cut-off parameter $\Delta E$. This can be
overcome by including emission of hard photons, with energy
$E_\gamma > \Delta E$
\begin{eqnarray}
\Gamma_\mathrm{brems} = \Gamma_\mathrm{soft}(\Delta E) +
\Gamma_\mathrm{hard}(\Delta E),
\end{eqnarray}
where $\Gamma_\mathrm{brems}$ is the total contribution to the decay
width due to photon emission.

To separate QED and SUSY corrections in the decay width we follow
the conventions of the Supersymmetry Parameter Analysis
(SPA)~\cite{spa,Drees:2006um}. From the sum of virtual and soft
photon terms
\begin{eqnarray}
\Gamma_\mathrm{virt}+\Gamma_\mathrm{soft} =
\Gamma_\mathrm{logs}+\Gamma_\mathrm{SUSY}
\end{eqnarray}
we take $\Gamma_\mathrm{logs}$ which contains potentially large
logarithms depending on small lepton mass $m_\ell$ and cut-off
energy $\Delta E$. The remaining part $\Gamma_\mathrm{SUSY}$ is IR
and UV finite, and free from $\Delta E$. We can now define the QED
correction as
\begin{eqnarray}
\Gamma_\mathrm{QED}=\Gamma_\mathrm{logs} + \Gamma_\mathrm{hard}\: .
\end{eqnarray}
Using above definitions we can rewrite the complete one-loop decay
width as
\begin{eqnarray}
\Gamma_\mathrm{loop} &=& \Gamma_\mathrm{tree} + \Gamma_\mathrm{virt}
+
\Gamma_\mathrm{brems}\nonumber \\
&=& \Gamma_\mathrm{tree} + \Gamma_\mathrm{SUSY} +
\Gamma_\mathrm{QED}\: . \label{eq:corr_split}
\end{eqnarray}

\section{Numerical analysis}\label{sec:4}
\sloppypar For the numerical analysis we take a modified SPS1a'
point~\cite{spa}. The modification is needed since in the original
parameter set there are two-body decay channels open for the light
chargino. To close these decay modes we decreased $M_2$ parameter
and increased slepton masses. The resulting spectrum is given in
Tab.~\ref{tab:mass}.

\begin{table}
\caption{Masses of particles in the chosen scenario.}
\label{tab:mass}
\begin{center}
\begin{tabular}{cccccc}\hline\noalign{\smallskip}
particle & $\tilde{\chi}^{\pm}_1$ & ${\tilde{\chi}^0_1}$ &
${\tilde{e}_L}$ & ${\tilde{e}_R}$ & ${\tilde{\nu}_e}$\\  mass [GeV]
 & 165.3 & 97.9 & 287.9 & 221.9 & 276.6 \\
\noalign{\smallskip}\hline\noalign{\smallskip} particle
&$\tilde{\tau}_1 $ & $\tilde{\tau}_2$
& $\tilde{q}_L$ & $ \tilde{q}_R$ & $H^{\pm}$\\
mass [GeV] & 211.9 & 289.0 & 561.3 & 544.3 & 436.4 \\ \hline
\end{tabular}
\end{center}
\end{table}

Results for one-loop corrections to the decay width in the
three-body leptonic chargino decays have been shown in
Tab.~\ref{tab:widths}. In our case corrections are of the order
$\sim 5\%$, however in other scenarios they can reach $\sim
10\%$~\cite{Fujimoto:2007bn}. We also checked the results for the
scenario introduced in~\cite{Fujimoto:2007bn} and found good
agreement\footnote{We note that there is an evident misprint in the
third column of Tab.~VII of Ref.~\cite{Fujimoto:2007bn}.}.

\begin{table}
\caption{One-loop corrected decay widths of $\tilde{\chi}^-_1$ in
keV.} \label{tab:widths}
\begin{center}
\begin{tabular}{lcc}
\hline\noalign{\smallskip}
decay mode & tree-level width &  one-loop width   \\
\noalign{\smallskip}\hline\noalign{\smallskip} $e^- \bar{\nu}_e
\tilde{\chi}_1^0$ & $4.18$
& $4.38$  \\
$\mu^- \bar{\nu}_{\mu} \tilde{\chi}_1^0$ & $4.18$ & $4.38$ \\
$\tau^- \bar{\nu}_{\tau}\tilde{\chi}_1^0$ & $4.38$  & 4.61  \\
\noalign{\smallskip}\hline
\end{tabular}
\end{center}
\end{table}

\begin{figure*}\sidecaption
\includegraphics[scale=0.4]{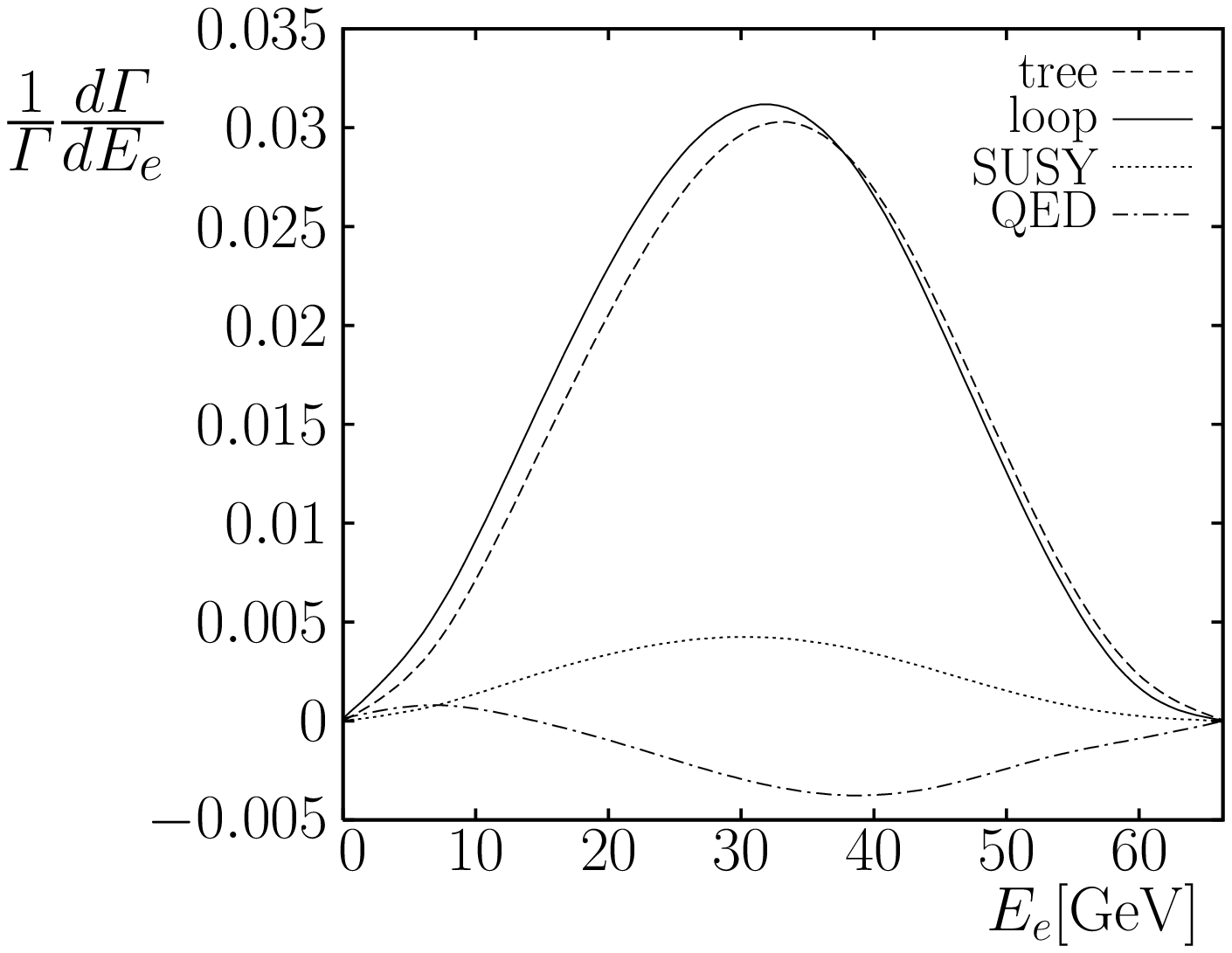}\hskip 0.5cm
\includegraphics[scale=0.4]{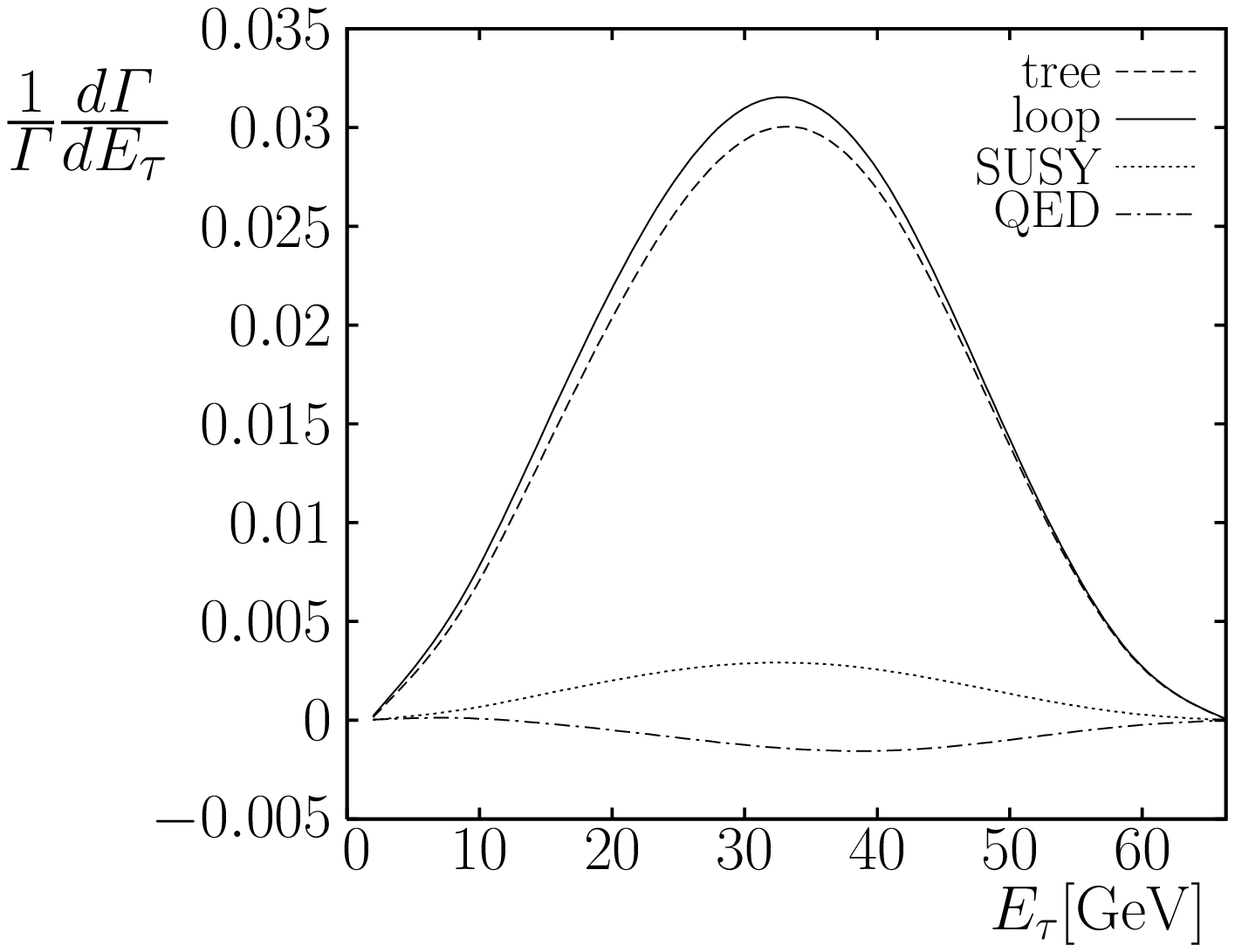}
\caption{One-loop corrected energy distributions for electron (left)
and $\tau$ (right) in the decay~(\ref{eq:decay}) together with
various contributions as in
Eq.~(\ref{eq:corr_split}).\label{fig:energy}}
\end{figure*}

\begin{figure}
\begin{center}
\includegraphics[scale=0.4]{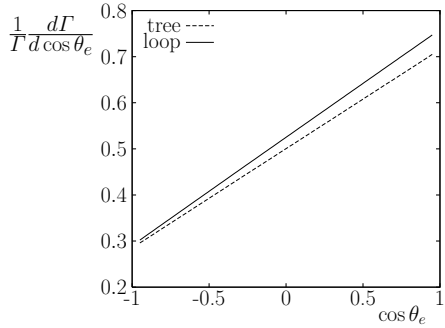}
\caption{One-loop corrected electron angular distribution with
respect to the chargino polarization vector in the
decay~(\ref{eq:decay}).\label{fig:angular}}
\end{center}
\end{figure}

In Fig.~\ref{fig:energy} we show the energy distributions of
electron and $\tau$ in the decay~(\ref{eq:decay}) at the one-loop
level. We also show separately the impact of QED and genuine SUSY
corrections, Eq.~(\ref{eq:corr_split}), for these distributions. As
can be seen, the shape of QED corrections is different for electron
and $\tau$ due to the difference in their masses. This difference
results in a shift of one-loop corrected energy distribution for
electrons towards lower energies. We also note that corrections for
electrons are in principle larger. Generally both QED and SUSY
corrections can be of the order $\sim 10\%$ but due to the opposite
relative sign we observe a partial cancelation of these
contributions. In Fig.~\ref{fig:angular} we show the electron
angular distribution with respect to the chargino polarization
vector. For some values of $\cos\theta_e$ the correction can reach
10\%.

\begin{figure}
\begin{center}
\includegraphics[scale=0.4]{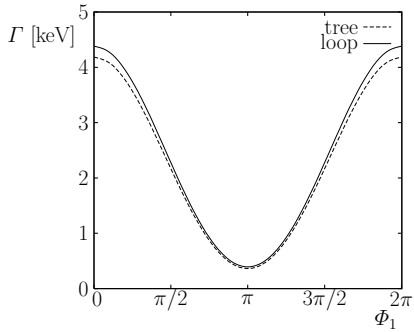}
\caption{Tree-level and one-loop corrected decay width for
$\tilde{\chi}^-_1 \to \tilde{\chi}_1^0 e^- \bar{\nu}$ as a function
of the phase of $M_1$ parameter.\label{fig:m1}}
\end{center}
\end{figure}

Figure~\ref{fig:m1} shows the dependence of the one-loop corrected
decay width $\tilde{\chi}^-_1 \to \tilde{\chi}_1^0 e^- \bar{\nu}$ on
the CP phase of the bino mass parameter $M_1$, which enters in the
neutralino couplings. This parameter has the strong influence on
$\Gamma$ and can change it by an order of magnitude. Although the
decay width is not a CP-odd observable this feature can provide some
information on the CP phase in the neutralino sector due to its
influence on the branching fractions of light chargino decay modes.

\section{Summary and outlook\label{sec:5}}
We have calculated the one-loop corrections to the three-body
leptonic chargino decays in the complex MSSM. This corrections may
turn out to be important for precision physics at the future linear
collider. The next step will be the inclusion of the hadronic decay
modes and incorporation of one-loop corrections to the full
production-decay process.

\begin{acknowledgement}
I would like to thank J.~Kalinowski for many inspiring discussions.
The author is supported by the Polish Ministry of Science and Higher
Education Grant No.~1~P03B~108~30.
\end{acknowledgement}

\end{document}